\documentclass[twocolumn,showpacs,preprintnumbers]{revtex4}
\usepackage{graphicx}
\usepackage{dcolumn}
\usepackage{bm}
\begin{document}
\newcommand{\ds}{\displaystyle}
\newcommand{\be}{\begin{equation}}
\newcommand{\en}{\end{equation}}
\newcommand{\bea}{\begin{eqnarray}}
\newcommand{\ena}{\end{eqnarray}}
\title{Two-fluid evolving Lorentzian wormholes}
\author{Mauricio Cataldo}
\altaffiliation{mcataldo@ubiobio.cl} \affiliation{Departamento de
F\'\i sica, Facultad de Ciencias, Universidad del B\'\i o-B\'\i o,
Avenida Collao 1202, Casilla 5-C, Concepci\'on, Chile.}
\author{Sergio del Campo}
\altaffiliation{sdelcamp@ucv.cl} \affiliation{Instituto de F\'\i
sica, Facultad de Ciencias, Pontificia Universidad Cat\'olica de
Valpara\'\i so, \\ Avenida Brasil 2950, Valpara\'\i so, Chile.}
\date{\today}
\begin{abstract}
We investigate the evolution of a family of wormholes sustained by
two matter components: one with homogeneous and isotropic properties
$\rho(t)$ and another inhomogeneous and anisotropic
$\rho_{in}(t,r)$. The rate of expansion of these evolving wormholes
is only determined by the isotropic and homogeneous matter component
$\rho(t)$. Particularly, we consider a family of exact two-fluid
evolving wormholes expanding with constant velocity and satisfying
the dominant and the strong energy conditions in the whole
spacetime. In general, for the case of vanishing isotropic fluid
$\rho(t)$ and cosmological constant $\Lambda$ the space expands with
constant velocity, and for $\rho(t)=0$ and $\Lambda \neq 0$ the rate
of expansion is determined by the cosmological constant. The
considered here two-fluid evolving wormholes are a generalization of
single fluid models discussed in previous works of the present
authors [Phys.\ Rev.\  D {\bf 78}, 104006 (2008); Phys.\ Rev.\  D
{\bf 79}, 024005 (2009)].

\vspace{0.5cm} 
\end{abstract}
\smallskip
\maketitle 
\section{Introduction}
Wormhole spacetimes have become one of the most popular and
intensively studied topics in general relativity. Throughout the
last decades there has been an accumulating volume of works on the
analytic wormhole geometries. The various approaches include both
static~\cite{StaticWH} and evolving relativistic
versions~\cite{EvolvingWH}. They principally consider static
wormhole spacetimes sustained by a single fluid component which
requires the violation of the null energy condition (NEC), and the
interest has been focused on traversable wormholes, which have no
horizons, allowing two-way passage through them. These hypothetical
tunnels in spacetime allow effective superluminal travels, although
the speed of light is not locally surpassed~\cite{LoboVisser}.

It is interesting to note that for constructing wormhole geometries
in general is adopted the reverse approach for solving the Einstein
field equations. This means that one first fixes the form of the
spacetime metric (such as the redshift and shape functions) and
then, by computing the field equations, one finds the
energy-momentum tensor components needed to support such a spacetime
geometry. The obtained in such a way stress components automatically
satisfy local conservation equations, by virtue of the Bianchi
identities. This reverse method helps us to find that a static
traversable wormhole violates the NEC~\cite{Morris,Visser}, thus in
general relativity an exotic type of matter is required for
sustaining a static traversable wormhole. It is interesting to note
that there are explicit static wormholes solutions respecting the
energy conditions in the whole spacetime in Einstein-Gauss-Bonnet
gravity~\cite{MaedaNozawa}. Notice also that in higher dimensions,
the presence of terms with higher powers in the curvature provided
by certain class of Lovelock theories, allows to remove the
possibility of violating energy conditions even in vacuum, since the
whole spacetime is devoid of any kind of stress-energy
tensor~\cite{Troncoso}.

However, it is well known that in Einstein gravity there are
nonstatic Lorentzian wormholes which do not require WEC violating
matter to sustain them. Such wormholes may exist for arbitrarily
small or large intervals of time~\cite{KarSahdev}.

On the purely gravitational side, most of the efforts are directed
to study Lorentzian wormholes sustained by a single exotic fluid in
classical general relativity. However, one can consider also
gravitational configurations filled with two or more
fluids~\cite{Pavon}. For example, in cosmology such two-fluid models
are widely considered today in order to explain the observed
accelerated expansion of the Universe~\cite{Pavon2}.

In this paper we shall study evolving wormholes sustained by two
fluids: one with homogeneous and isotropic properties $\rho(t)$ and
another inhomogeneous and anisotropic $\rho_{in}(t,r)$. The
theoretical construction of these wormholes will be performed by
imposing conditions on the stress-energy tensor threading the
evolving wormhole geometry. Specifically, we shall consider the
radial and tangential pressures of the inhomogeneous and anisotropic
matter to obey barotropic equations of state with constant state
parameters. On the other hand, the homogeneous and isotropic fluid
is taken to be that of a perfect fluid described by the
energy-momentum tensor
\begin{eqnarray}\label{TT}
T_{\alpha \beta}=(\rho+p) u_{\alpha}u_{\beta}-p g_{\alpha \beta},
\end{eqnarray}
where $u_{\alpha}$ is the four-velocity of the fluid, $\rho$ and $p$
are the energy density and the pressure of the cosmic fluid
respectively

We shall suppose that the dynamics of the gravitational fields is
governed by Einstein field equations
\begin{eqnarray*}
R_{\alpha \beta}-\frac{R}{2} g_{\alpha \beta}=-\kappa T_{\alpha
\beta}-\Lambda g_{\alpha \beta},
\end{eqnarray*}
where $\kappa=8 \pi G$ and $\Lambda$ is the cosmological constant,
and the evolving wormhole metric will be given by
\begin{eqnarray}\label{evolving wormhole15}
ds^2=-e^{2\Phi(t,r)}dt^2+  a(t)^2 \left(
\frac{dr^2}{1-\frac{b(r)}{r}}+r^2 d \Omega^2 \right),
\end{eqnarray}
where $\Phi(t,r)$ is the redshift function, $a(t)$ is the scale
factor of the wormhole universe, $b(r)$ is the shape function and $d
\Omega^2=d\theta^2+sin^2 \theta d \varphi^2$. Note that the
essential characteristics of a wormhole geometry are encoded in the
spacelike section of the metric~(\ref{evolving wormhole15}). It is
clear that this metric becomes a static wormhole if $a(t)
\rightarrow const$ and $\Phi(t,r)=\Phi(r)$ and, as $b(r) \rightarrow
0$ and $\Phi(t,r) \rightarrow 0$ it becomes a flat
Friedmann-Robertson-Walker metric.

The organization of the paper is as follows: In Sec. II we present
the dynamical field equations for wormhole models with a matter
source composed of an ideal isotropic cosmic fluid and an
anisotropic and inhomogeneous one. In Sec. III some aspects of the
geometry of the general solution are discussed. In Sec. IV expanding
wormholes are discussed, and in Sec. V we conclude with some
remarks.

\section{Field equations}

Let us now consider the dynamical field equations describing
evolving wormhole models~(\ref{evolving wormhole15}). We shall be
interested in studying wormhole scenarios filled with two fluids
$\rho=\rho(t)$ and $\rho_{_{in}}=\rho_{_{in}}(t,r)$, where the first
cosmic fluid always remains homogeneous and isotropic and the other
component is in general an inhomogeneous and anisotropic fluid.
Since we have a spherically symmetric space-time, the cosmic fluid
$\rho_{_{in}}(t,r)$ in general may have anisotropic pressures, which
we shall define as $p_{r}(t,r)$ and $p_{l}(t,r)$ for the radial and
lateral components respectively. If $p_{r}(t,r)=p_{l}(t,r)$ we have
an isotropic inhomogeneous pressure.

Thus, for spherically symmetric spacetimes written in comoving
coordinates~(\ref{evolving wormhole15}) and filled with these two
kinds of cosmic fluids, the Einstein field equations may be written
in the following form:
\begin{eqnarray}\label{00}
3 e^{-2\phi(t,r)} H^2+\frac{b^{\prime}}{a^2 r^2}=\kappa \rho_{_{in}}(t,r)+\kappa \rho(t)+\Lambda,  \\
\label{rr} - e^{-2\phi(t,r)}\, \left( 2\frac{\ddot{a}}{a}+ H^2
\right)- \frac{b}{a^2 r^3} + 2 e^{-2\phi(t,r)} H \frac{\partial
\phi}{\partial t}+ \\
\nonumber \frac{2}{r^2 a^2} (r-b ) \frac{\partial \phi}{\partial r}=\kappa p_r(t,r)+\kappa p(t)-\Lambda, \\
\label{thetatheta} -e^{-\phi(t,r)} \left(2 \frac{\ddot{a}}{a}+ H^2
\right) + \frac{b-r b^{\prime}} {2 a^2 r^3}+ \\ \nonumber 2
e^{-\phi(t,r)} H \frac{\partial \phi}{\partial t} +
\frac{1}{2a^2r^2} (2r-b-rb^{\prime}) \frac{\partial \phi}{\partial
r} + \\ \nonumber \frac{1}{a^2r} (r-b)\left( \left(\frac{\partial
\phi}{\partial r} \right)^2+\frac{\partial^2 \phi}{\partial
r^2}\right)=\kappa
p_{_l}(t,r)+\kappa p(t) -\Lambda,  \\
2 e^{-\phi(t,r)} \sqrt{\frac{r-b(r)}{r}} \,  \frac{\partial
\phi}{\partial r} \, \dot{a}=0, \label{thetaphi}
\end{eqnarray}
where it was assumed that the 4-velocity of both fluids is the
timelike vector $u^{\alpha}=(e^{-\phi},0,0,0)$, $\kappa=8 \pi G$,
$H=\dot{a}/a$, and an overdot and a prime denote differentiation
$d/dt$ and $d/dr$ respectively.

In attempting to find solutions to the field equations we first note
that Eq.~(\ref{thetaphi}) gives some constraints on relevant metric
functions which separate the wormhole solutions into two branches:
one static branch given by the condition $\dot{a}=0$ and another
non-static branch for $\partial \phi/\partial r=0$.

In what follows we shall restrict our discussion to nonstatic
branch. The condition $\partial \phi/\partial r=0$ implies that the
redshift function can only be a function of $t$, i.e.
$\phi(t,r)=f(t)$ so, without any loss of generality, by rescaling
the time coordinate we can set $\phi(t,r)=0$. Thus we shall look for
the solutions to Einstein equations described by the metric form
\begin{eqnarray}\label{evolving wormhole15}
ds^2=-dt^2+  a(t)^2 \left( \frac{dr^2}{1-\frac{b(r)}{r}}+r^2 d
\Omega^2 \right),
\end{eqnarray}
so we have to put $\Phi=0$ into the
Eqs.~(\ref{00})-(\ref{thetaphi}). As we shall see below this will
imply that the anisotropic and inhomogeneous matter component
$\rho_{_{in}}(t,r)$ cannot be isotropic.

Notice that the field equations~(\ref{00})-(\ref{thetaphi})
generalize the Einstein equations considered in
Refs.~\cite{Cataldo1} and~\cite{Cataldo2}. Thus, in order to quickly
solve the field equations we shall use the conservation equations
$T^\mu_{\,\,\,\,\nu;\mu}=0$. By supposing that each fluid satisfies
the standard conservation equation separately we obtain
\begin{eqnarray}
\frac{\partial \rho}{\partial t}+3H (\rho+p)=0, \label{FRWCE1}\\
\frac{\partial \rho_{_{in}}}{\partial t}+H (3 \rho_{_{in}}+p_r+2p_{_l})=0, \label{FRWCE2} \\
\frac{2(p_{_l}-p_r)}{r}=\frac{\partial p_r}{\partial r},
\label{FRWCE3}
\end{eqnarray}
where Eq.~(\ref{FRWCE1}) states the conservation of the isotropic
and homogeneous component and Eqs.~(\ref{FRWCE2}) and~(\ref{FRWCE3})
are valid for the anisotropic and inhomogeneous cosmic fluid and may
be interpreted as the conservation equation and the relativistic
Euler equation (or the hydrostatic equation for equilibrium for the
anisotropic matter supporting the gravitational configuration)
respectively.

Notice that from equations~(\ref{FRWCE2}) and~(\ref{FRWCE3}) we see
that for an isotropic but still inhomogeneous matter component
$\rho_{_{in}}$, i.e. $p_{_l}=p_r=p_{_{in}}$, we have to require
$\partial p_r/\partial r=0$, so the pressure will depend only on
time $t$, obtaining the standard cosmological conservation equation
$\dot{\rho}_{_{in}}(t)+3 H [\rho_{_{in}}(t)+p_{_{in}}(t)]=0$. Thus
in this case we have a noninteracting superposition of two
homogeneous and isotropic fluids $\rho$ and $\rho_{_{in}}$. This
leads us to conclude that if we want to study evolving wormholes
filled with a mixture of homogeneous, isotropic fluid $\rho(t)$ and
an inhomogeneous fluid $\rho_{_{in}}(t,r)$, we must consider only
anisotropic matter component $\rho_{_{in}}(t,r)$ with $ p_r \neq
p_{_l}$.

For solving the field equations of the considered gravitational
configuration we shall consider the following anzats: we shall
require that the radial and the lateral pressures have barotropic
equations of state. Thus we shall write for them
\begin{eqnarray} \label{FRWC00}
p_r(t,r)=\omega_r \, \rho_{_{in}}(t,r), \nonumber \\
p_{_l}(t,r)=\omega_{_l} \, \rho_{_{in}}(t,r),
\end{eqnarray}
where $\omega_r$ and $\omega_{_l}$ are constant state parameters
(note that in this case we have that
$p_{_{\theta}}=p_{_{\phi}}=p_{_l}$ due to the spherical symmetry).

In the following, for solving the field equations, we shall require
that $p_r(t,r) \neq p_{_l}(t,r)$. By taking into account
Eq.~(\ref{FRWC00}), from Eq.~(\ref{FRWCE3}) we get
\begin{equation}
\rho_{_{in}}(t,r)=F(t) \, r^{2(\omega_{_l}-\omega_r)/\omega_r},
\end{equation}
where $F(t)$ is an integration function, and introducing this
expression into Eq.~(\ref{FRWCE2}) we have for the energy density of
the anisotropic matter
\begin{equation}\label{rhoC}
\rho_{_{in}}(t,r)=\frac{C \, r^{2(\omega_{_l}-\omega_r)/\omega_r}}{
a^{3+\omega_r+2\omega_{_l}}},
\end{equation}
where $C$ is an integration constant.

Now, by subtracting Eqs.~(\ref{rr}) and~(\ref{thetatheta}), and
using the full energy density~(\ref{rhoC}), we obtain the
differential equation
\begin{equation}
\frac{\kappa (\omega_{_l}-\omega_r)\, C \,
r^{2(\omega_{_l}-\omega_r)/\omega_r}}{a^{(3+\omega_r+2\omega_{_l})}}=\frac{3b-r
b^{\prime}}{2a^2r^3}.
\end{equation}
It is straightforward to see that in order to have a solution for
the shape function $b=b(r)$ we must impose the constraint
\begin{equation}\label{constraint15}
\omega_r+2\omega_{_l}+1=0
\end{equation}
on the state parameters $\omega_r$ and $\omega_{_l}$, thus obtaining
for the shape function
\begin{equation}\label{br}
b(r)=C_3 r^3-\kappa \, C \, \omega_r \, r^{-1/\omega_r},
\end{equation}
where $C_3$ is a new integration constant. Notice that the
constraint~(\ref{constraint15}) implies that the radial and
tangential pressures are given by
\begin{equation}\label{pr pl}
p_r=\omega_r \rho_{_{in}}, \,\,\, p_{_l}= -\frac{1}{2} \,
(1+\omega_r) \rho_{_{in}},
\end{equation}
so the energy density and pressures satisfy the following relation:
\begin{equation}
\rho_{_{in}}+p_r+2 p_{_l}=0.
\end{equation}
This equation implies that the inhomogeneous and anisotropic
component satisfies the strong energy condition. In this case for
$\omega_{_r} \leq -1$, $-1 \leq  \omega_{_r} \leq  1$ and
$\omega_{_r} \geq 1$ we have that $\omega_{_l} \geq 0$, $-1 \leq
\omega_{_l} \leq 0$ and $\omega_{_l} \leq  -1$ respectively.

Now, from Eqs.~(\ref{00}),~(\ref{rhoC}),~(\ref{br}) and taking into
account the constraint~(\ref{constraint15}) we obtain the following
master equation for the scale factor:
\begin{equation}\label{scf}
3 H^2=-\frac{3C_3}{a^2}+ \kappa \rho + \Lambda.
\end{equation}

Note that by taking into account the metric~(\ref{evolving
wormhole15}) we conclude that $C_3$ may be absorbed by rescaling the
$r$-coordinate as follows: $C_3=1$ for $C_3>0$ and $C_3=-1$ for
$C_3<0$, so without any loss of generality we can identify it with
the spatial curvature parameter $k$ by putting $C_3=k$, with
$k=-1,0,1$. Thus the master equation~(\ref{scf})) may be written in
the form
\begin{eqnarray} \label{friedmann equation}
3 H^2+\frac{3k}{a^2} = \kappa \rho(t)+\Lambda.
\end{eqnarray}

Summarizing, we have shown that for the gravitational configuration
\begin{eqnarray}\label{INHOMOGENEOUS FRW}
ds^2=dt^2-a(t)^2  \times \nonumber \\ \left( \frac{dr^2}{1-k
r^2+\kappa \, C \, \omega_r
\, r^{-1-1/\omega_r}}+ r^2 (d\theta^2+sin^2 \theta d \varphi^2)\right), \nonumber \\
\end{eqnarray}
filled with the inhomogeneous cosmic fluid
\begin{equation}\label{rhoC Final}
\rho_{_{in}}(t,r)=\frac{C \, r^{-3-1/\omega_r}}{ a^{2}(t)},
\end{equation}
(with anisotropic pressures $p_r=\omega_r \rho_{_{in}}$ and
$p_l=-\frac{1}{2}(1+\omega_r) \rho_{_{in}}$) and another
noninteracting arbitrary homogeneous and isotropic $\rho(t)$, the
evolution of the scale factor $a(t)$ is governed by the standard
Friedmann equation~(\ref{friedmann equation}) and the conservation
equation~(\ref{FRWCE1}).

In conclusion, the rate of expansion of these evolving wormholes is
only determined by the matter component $\rho(t)$ which may be in
principle an ideal barotropic fluid, a scalar field or any other
isotropic and homogeneous cosmic fluid considered in literature.
Notice that if $\rho(t)=\Lambda=0$ the space expands with constant
velocity~\cite{Cataldo1}, and for $\rho(t)=0$ and $\Lambda \neq 0$
the rate of expansion is determined by the cosmological
constant~\cite{Cataldo2}.

It is worth noticing that for $\omega_r=-1/3$ we have a limiting
case since the 3-space becomes isotropic and homogeneous and the
anisotropic matter behaves as an ideal string gas (i.e.
$p_r=p_{_l}=-\rho_{_{in}}(t)/3$). From Eq.~(\ref{rhoC Final}) we
conclude that in this case the matter component behaves as
$\rho_{_{in}}=C/a^2(t)$, implying that we have a FRW cosmology
filled with a mixture of a curvature fluid with a cosmic fluid
$\rho(t)$.

\section{Some remarks on the geometry of the space-time}
It is clear that the gravitational configuration~(\ref{INHOMOGENEOUS
FRW}) is sustained via a matter source made of the inhomogeneous and
anisotropic cosmic fluid~(\ref{rhoC Final}). Let us point out some
properties of the discussed geometry. In general the
metric~(\ref{INHOMOGENEOUS FRW}) is not conformally flat since the
Weyl tensor does not vanish for this metric, except for $C=0$ or $C
\neq 0$ and $\omega_r=-1/3$. On the other hand, this
acceleration-free space-time ($g_{tt}=1$) is characterized by zero
anisotropic stress $\sigma_{\alpha \beta}(t,r)$ and zero heat-flux
vector $q_{\alpha}$.

Note that the metric~(\ref{INHOMOGENEOUS FRW}) is conformal to the
following static metric:
\begin{eqnarray}\label{static metric}
ds^2=d\tau^2- \left( \frac{dr^2}{1-k r^2+\kappa \, C \, \omega_r \,
r^{-1-1/\omega_r}}+ r^2 d \Omega^2 \right), \hspace{.3cm}
\end{eqnarray}
where $d \Omega^2=d\theta^2+sin^2 \theta d \varphi^2$ and $\tau=\int
dt/a(t)$ is the conformal time. In general the component
$g_{rr}^{-1}$ of the metric~(\ref{static metric})  may be valid for
all $r>0$ or vanish for some value of the radial coordinate $r_0
>0$; however this does not mean that this space-time contains an
event horizon at $r_0$ since $g_{\tau \tau}=1$. So in principle, for
some sets of the model parameters, this space-time may contain a
naked singularity at $r=0$ which may be observable from the outside.

In general this geometry admits three-dimensional slices
$t=t_0=const$ with a variable curvature. In this case the
3-curvature may be written as
\begin{eqnarray}\label{curvature}
{}^3R=-a_0^{-2}\left(6k+2\kappa C  r^{-3-\frac{1}{\omega_r}}\right),
\end{eqnarray}
where $a_0=a(t_0)$. If $\omega_r<-1/3$ or $\omega_r>0$ these slices
are asymptotically flat for $k=0$, asymptotically de-Sitter for
$k=1$ or asymptotically anti de-Sitter for $k=-1$. Note that for
these ranges of $\omega_r$ there may arise a naked singularity at
$r=0$, and on the other hand the energy density of the inhomogeneous
matter component vanishes as $r \rightarrow \infty$ since from
Eq.~(\ref{rhoC Final}) we have that $\rho_{_{in}}(t_0,r) \rightarrow
0$. It is remarkable that the $3$-dimensional slices $t=t_0$ of
metric~(\ref{INHOMOGENEOUS FRW}) include as a particular case the
$3$-dimensional slices $t=t_0$ of the Kottler metric
\begin{eqnarray}\label{Kottler}
ds^2=\left(1-\frac{2M}{r}-\frac{\Lambda}{3} r^2 \right) dt^2-
\frac{dr^2}{1-\frac{2M}{r}-\frac{\Lambda}{3} r^2}+ r^2 d \Omega^2,
\nonumber \\
\end{eqnarray}
which includes the de-Sitter ($\Lambda>0$) and anti de-Sitter
($\Lambda<0$) space-times. The $3$-curvature of its slices $t=t_0$
is given by ${}^3R=-2\Lambda$, so they are also spaces of constant
curvature. By comparing metrics~(\ref{INHOMOGENEOUS FRW})
and~(\ref{Kottler}) we conclude that Kottler slices $t=t_0$ are
obtained from slices of metric~(\ref{INHOMOGENEOUS FRW}) by putting
$a_0^{-2}k=\Lambda/3$, $\kappa C \omega_r=-2M$ and $\omega_r
\rightarrow \pm \infty$. From Eq.~(\ref{curvature}) we see that
\begin{eqnarray*}
{}^3R a_0^{2}=-6k-2\kappa C r^{-3-\frac{1}{\omega_r}} \equiv \nonumber \\
-6k-2\frac{\kappa C \omega_r}{\omega_r} \, r^{-3-\frac{1}{\omega_r}}
\longrightarrow  -6k,
\end{eqnarray*}
for $\omega_r \rightarrow \pm \infty$, thus effectively we have in
this case a space of constant curvature. It can be shown that for
this limit the inhomogeneous energy density~(\ref{rhoC Final})
vanishes since
\begin{eqnarray}
\rho_{_{in}}(t,r)=\frac{C \, r^{-3-1/\omega_r}}{ a^{2}(t)} \equiv
\frac{C \omega_r \, r^{-3-1/\omega_r}}{ \omega_r a^{2}(t)}
\longrightarrow 0
\end{eqnarray}
for $\omega_r \rightarrow \pm \infty$, while the anisotropic
pressures take the forms $p_r(t,r) \rightarrow  \frac{\omega_r
C}{a^2(t) r^3}$ and $p_{_l}(t,r) \rightarrow \frac{\omega_r
C}{2a^2(t) r^3}$. Unfortunately this model with a vanishing
inhomogeneous and anisotropic energy density $\rho_{_{in}}(t,r)$ and
non-vanishing pressures $p_r(t,r)$ and $p_{_l}(t,r)$ is
non-physical, so we rule it out from consideration.

If $-1/3<\omega_r<0$ the curvature increases with radius and as $r
\rightarrow \infty$ we have the asymptotic forms ${}^3R \sim
r^\alpha$, $\rho_{_{in}} \sim r^\alpha$ and $f^2(r) \sim r^\beta$
with $\alpha>0$ and $\beta>2$ respectively. These gravitational
configurations do not contain any singularity at $r=0$ for slices
$t=t_0$.

\section{Expanding Wormhole Universes}

Now we shall study gravitational configurations where the considered
solution represents an expanding wormhole
geometry~\cite{Morris,Visser}. Before treating Lorentzian wormhole
geometries in more detail, let us note again that the metric
ansatz~(\ref{evolving wormhole15}) provides an explicit class of
dynamic wormholes that generalize the static, spherically symmetric
ones first considered by Morris and Thorne~\cite{Morris}. Several
other aspects of static and evolving wormhole spacetimes are
analyzed in Refs.~\cite{StaticWH} and~\cite{EvolvingWH}
respectively.

In order to have a wormhole geometry the functions $\Phi(r)$ and
$b(r)$ must satisfy some constraints defined by the authors of
Ref.~\cite{Morris}. However, these authors originally were
interested only in wormhole geometries featuring two asymptotically
flat regions connected by a bridge. In our case we are interested in
a more general asymptotic behavior. Due to the presence of the
constant $k=-1,0,1$; besides the asymptotically flat wormholes
($k=0$) we may have anti--de Sitter ($k=-1$) asymptotic wormholes
which also may be of particular interest~\cite{Barcelo}. Thus the
main constraints may be defined as follows:

Constraint 1: A no--horizon condition, i.e. $e^{\Phi(r)}$ is finite
throughout the space--time in order to ensure the absence of
horizons and singularities.

Constraint 2: The shape function $b(r)$ must obey at the throat $r =
r_0$ the following condition: $b(r_0) = r_0$, being $r_0$ the
minimum value of the $r$--coordinate. In other words
$g^{-1}_{rr}(r_0)=0$.

Constraint 3: Finiteness of the proper radial distance
\begin{eqnarray}
l(r)=\pm \int^r_{r_0} \frac{dr}{\sqrt{1-b(r)/r}}
\end{eqnarray}
for $r \geq r_0$ throughout the space--time. The $\pm$ signs refer
to the two regions which are connected by the wormhole.

Let us now consider the possibility of having a wormhole geometry.
From the metric~(\ref{INHOMOGENEOUS FRW}) we conclude that
$e^{\Phi(r)}=1$, thus the first constraint is automatically
fulfilled. It must be remarked that, for a general dynamical
wormhole, the definition of the location of a wormhole throat is not
straightforward. In this case, the position of the wormhole throat
depends on the time-slicing, and for a time-dependent wormhole it
may not be possible to locate the entire throat within one time
slice, as the dynamic throat is an extended object in
spacetime~\cite{Hochberg}. In our case, the evolving
metric~(\ref{INHOMOGENEOUS FRW}) is conformal to the static
spacetime~(\ref{static metric}), which represents a static wormhole
for $\omega_r < -1$ or $\omega_r
>0$, and then we always may determine the location of the wormhole
throat on the hypersurface with $t=t_0=const$. In general, for
time-dependent spherically symmetric wormhole spacetimes,
alternative definitions of a wormhole throat are required, equally
valid for static as well as for dynamical wormholes. Several
different definitions have been given in
Refs.~\cite{MaedaNozawa,Hochberg,Maeda,Hayward}. For example
Hochberg and Visser~\cite{Hochberg} and Hayward~\cite{Hayward} have
introduced two independent quasilocal definitions of a throat for
dynamical wormholes.  These authors do not consider global
properties of the wormholes, i.e. they make no assumptions about
symmetries, asymptotic flatness, topology, etc., and the wormhole
throat is a two-dimensional surface of nonvanishing minimal area on
a null hypersurface. On the other hand, in the Ref.~\cite{Maeda} the
authors have defined a wormhole throat quasilocally in terms of a
surface of nonvanishing minimal area on a spacelike hypersurface.
Some properties of these three definitions are compared
in~\cite{Maeda}.

From the second constraint, $g^{-1}_{rr}(r_0)=0$, we can find the
minimum value $r_0$ of the radial coordinate where the wormhole
throat must be located. From this throat condition, and by taking
into account the metric~(\ref{INHOMOGENEOUS FRW}), we obtain for the
integration constant
\begin{eqnarray}
C=\frac{(kr^2_0-1)}{\kappa \omega_r} \, r_0^{(1+\omega_r)/\omega_r},
\end{eqnarray}
yielding for the shape function $b(r)$ and the metric component
$g_{rr}$
\begin{eqnarray}\label{whevolving}
b(r)=r_0\left(\frac{r}{r_0}\right)^{-1/\omega_r}
\,\,\,\,\,\,\,\,\,\,\,\,\,\,\,\,\,\,\,\,\,\,\,\,\,\,\,\,\,\,
\nonumber
\\ +kr^3_0 \left(\frac{r}{r_0}\right)^3
\left(1-\left(\frac{r}{r_0}\right)^{-(1+3\omega_r)/\omega_r}\right),
\nonumber \\ a^2(t) g_{rr}^{-1}= 1-
\left(\frac{r}{r_0}\right)^{-(1+\omega_r)/\omega_r}
\,\,\,\,\,\,\,\,\,\,\,\,\,\,\,\,\,\,\,\,\,\,\,\,\,\,\,\,\,\,
\nonumber
\\ -kr^2_0 \left(\frac{r}{r_0}\right)^2
\left(1-\left(\frac{r}{r_0}\right)^{-(1+3\omega_r)/\omega_r}\right),
\end{eqnarray}
respectively.

It is easy to verify that the wormhole throat is located at $r_0$
since $b(r_0)=r_0$. In this case the energy density of the matter
threading the wormhole takes the following form:
\begin{eqnarray}\label{rrr}
\kappa \rho_{_{in}}(t,r)=\frac{kr^2_0-1}{r^2_0 \omega_r a(t)^2}
\left(\frac{r}{r_0}\right)^{-(1+3\omega_r)/\omega_r}.
\end{eqnarray}
Clearly, in order to have an evolving wormhole we must require
$\omega_r < -1$ or $\omega_r >0$ (in both of these cases, in the
$g_{rr}$ metric component, $(1+\omega_r)/\omega_r>0$ and
$(1+3\omega_r)/\omega_r>0$), implying that the inhomogeneous and
anisotropic cosmic fluid~(\ref{rrr}) can support the existence of
evolving wormholes. It can be shown that the form of the wormhole is
preserved during all evolution. Notice that for an anisotropic
matter with $\omega_r<-1$ we have that $\rho_{_{in}}(t,r)>0$, while
for an anisotropic matter with $\omega_r>0$ we have that
$\rho_{_{in}}(t,r)<0$. For $\rho(t)=\Lambda=0$ the shape of the
wormhole expands with constant velocity~\cite{Cataldo1}. In the
presence of a cosmological constant with $\rho_{_{in}}(t,r)=0$ the
wormhole configurations have an accelerated expansion
(contraction)~\cite{Cataldo2}. Other properties of such evolving
wormholes are discussed by authors of the
Ref.~\cite{Cataldo1,Cataldo2} and references cited therein.

As explicit examples of evolving wormholes, let us first consider
the case where the isotropic and homogeneous component is given by a
perfect fluid with the barotropic state equation $p(t)=\omega
\rho(t)$ and $k=\Lambda=0$. Thus the scale factor is given by
\begin{equation}\label{WH FRW}
a(t)=a_0 \, t^{2/3(1+\omega)}
\end{equation}
and the energy density by
\begin{equation}
\rho(t)={\frac {4}{3\kappa \left( 1+\omega \right) ^{2}\,{t}^{2}}},
\end{equation}
while the metric takes the form
\begin{eqnarray}\label{INHOMOGENEOUS FRW15}
ds^2=dt^2-a_0^2 \, t^{4/3(1+\omega)}  \times \nonumber \\ \left(
\frac{dr^2}{1-
\left(\frac{r}{r_0}\right)^{-(1+\omega_r)/\omega_r}}+ r^2 (d\theta^2+sin^2 \theta d \varphi^2)\right). \nonumber \\
\end{eqnarray}
In this case, the energy density of the anisotropic matter and its
pressure components are given by
\begin{equation}\label{WH FRW15}
\kappa \rho_{_{in}}(t,r)=-\frac{1}{\omega_r r^2_0 a_0^2 \,
t^{4/3(1+\omega)}}
\left(\frac{r}{r_0}\right)^{-(1+3\omega_r)/\omega_r}
\end{equation}
and~(\ref{pr pl}) respectively. Note that if $\omega_r<-1$ or
$\omega_r>0$ we have asymptotically flat FRW regions for $r
\longrightarrow \infty$ and $\omega>-1$. Let us suppose that the
isotropic fluid satisfies the dominant energy condition, i.e. $ |p|
\leq \rho$, $\rho \geq 0$. Thus if $-1 \leq \omega < -1/3$ the
expansion of the evolving wormhole is accelerated, i.e. both
universes and the throat of the wormhole are simultaneously
expanding with acceleration, while for $-1/3<\omega \leq1$ the
expansion is decelerated.

On the other hand, the total matter content is given by
$\rho_{_{total}}= \rho(t)+\rho_{_{in}}(t,r)$ and for any
$\omega_r<0$ we have that $\rho_{_{total}} \geq 0$. For $\omega_r>0$
we can have in general time intervals where the total energy is
positive or negative. For $\omega>-1/3$ the wormhole model starts
with a positive total energy density (since for a fixed $r=const$
the isotropic component dominates over the another one), then
decreases till zero at certain $t_{eq}$, and becomes negative for
$t> t_{eq}$. For $\omega < -1/3$ the total energy density starts
negative, then increases till zero at certain $t_{eq}$, and becomes
positive for $t> t_{eq}$.

The case $\omega=-1/3$ is more interesting, since it allows us to
consider evolving wormhole models satisfying the dominant energy
condition (DEC) in the whole spacetime. In this case the total
energy density and the corresponding total pressure components are
given by
\begin{eqnarray}
\rho_{_T}=\rho+\rho_{in}=\left ( 3-\frac{(r/r_0)^{-\frac{1+3
\omega_r}{\omega_r}}}{\omega_r r_0^2 a_0^2}  \right) t^{-2}, \nonumber \\
p_{_{r,T}}=-\frac{1}{3} \rho+\omega_r \rho_{in}, \nonumber \\
p_{_{l,T}}=-\frac{1}{3} \rho-\frac{1}{2}(1+\omega_r) \rho_{in}.
\end{eqnarray}
From this expressions we conclude that the total matter content
satisfies the strong energy condition since $\rho_{_T}+p_{_{r,T}}+2
p_{_{l,T}} \equiv 0$.

In order to fulfill the DEC we need to satisfy the following
conditions: $\rho_{_T} \geq 0$, $\rho_{_T} + p_{_{r,T}} \geq 0$,
$\rho_{_T} - p_{_{r,T}} \geq 0$, $\rho_{_T} + p_{_{l,T}} \geq 0 $
and $\rho_{_T} - p_{_{l,T}} \geq 0$. These conditions imply that the
following constraints must be satisfied:
\begin{eqnarray}\label{C15A}
\frac{1}{3 \omega_r } \leq a_0^2 r_0^2, \\
\frac{1+\omega_r}{2 \omega_r } \leq a_0^2 r_0^2, \\
\frac{1-\omega_r}{4 \omega_r } \leq a_0^2 r_0^2, \\
\frac{3+\omega_r}{8 \omega_r } \leq a_0^2 r_0^2. \label{C15B}
\end{eqnarray}
Hence, for any given $\omega_r > 0$, the DEC is fulfilled by
choosing the parameters $a_0^2$ and $r_0^2$ satisfying the
conditions~(\ref{C15A})-(\ref{C15B}). Note that in this case the
isotropic fluid gives a scale factor given by $a(t)=t$. Thus, the
class of analytic two-fluid evolving wormholes with $\omega=-1/3$
satisfies the dominant and strong energy conditions in the whole
spacetime and expands with constant velocity. It is interesting to
note that the violation or not of the NEC, at and near the throat,
of a dynamical wormhole, may be properly connected to the
generalizations of the flare-out condition for an arbitrary wormhole
discussed by above cited authors of the
Refs.~\cite{Hochberg,Maeda,Hayward}. In the definitions
of~\cite{Hochberg,Hayward} dynamical wormhole throats are trapping
horizons, i.e. hypersurfaces foliated by ``marginally trapped
surfaces", and the violation of the NEC is a generic property of
such wormhole throats. While in the Ref.~\cite{Maeda} a dynamical
spherically symmetric wormhole throat is defined as a ``trapped
sphere", and the NEC can still be satisfied for some wormhole
configurations.

As a second example, we shall consider an evolving wormhole with
$k=0$ and filled with a minimally coupled scalar field with the
exponential potential $V(\phi)=V_0 e^{-\lambda \phi}$, where
$\lambda$ and $V_0$ are constant parameters. In this case the energy
density and the pressure of the homogeneous and isotropic component
are given by $\rho_{\phi}=\dot{\phi}^2/2+V(\phi)$ and
$p_{\phi}=\dot{\phi}^2/2-V(\phi)$ respectively. Thus
Eqs.~(\ref{FRWCE1}) and~(\ref{friedmann equation}) imply that
\begin{eqnarray}
\ddot{\phi}(t) + 3H \dot{\phi} - \lambda V_0 e^{-\lambda \phi}=0, \\
3 H^2=\kappa \left(\frac{\dot{\phi}^2}{2}+V(\phi) \right).
\end{eqnarray}
A particular exact solution, describing the power-law expansion
\begin{eqnarray}\label{powerlaw}
a(t)=a_0 t^p
\end{eqnarray}
of the evolving wormhole, is given by
\begin{eqnarray}
\phi(t)=\frac{2}{\lambda} \ln \, t,
\end{eqnarray}
where
\begin{eqnarray}
\lambda^2=\frac{2 \kappa}{p}, V_0=\frac{p(3p-1)}{\kappa}.
\end{eqnarray}
The first relation implies that $p>0$, and from the second one we
have that $V_0<0$ if $0<p<1/3$ and $V_0>0$ if $p>1/3$ .

Finally, as a last example we shall consider evolving wormholes with
$k=0$ and filled with a tachyon field giving the power-law
expansion~(\ref{powerlaw}), where $p$ is a constant parameter. In
this case the energy density and the pressure of the homogeneous and
isotropic component are given by
$\rho_{_T}=\frac{V(\phi)}{\sqrt{1-\dot{\phi}^2}}$ and $p_{_T}=
-V(\phi) \sqrt{1-\dot{\phi}^2}$ respectively. Thus
Eqs.~(\ref{FRWCE1}) and~(\ref{friedmann equation}) imply that
\begin{eqnarray}
\frac{\ddot{\phi}(t)}{1-\dot{\phi}^2} + 3H \dot{\phi} +\frac{1}{V(\phi)} \frac{dV(\phi)}{d\phi}=0, \\
3 H^2=\frac{\kappa V(\phi)}{\sqrt{1-\dot{\phi}^2}}.
\end{eqnarray}
It can be shown that in order to have the power-law
expansion~(\ref{powerlaw}) the tachyon potential takes the form
$V(\phi)=\alpha \phi^{-2}$, where $\phi(t)=\phi_0 t$,
$\phi_0=\sqrt{2/3p}$ and $\kappa \alpha= 2p \sqrt{1-2/(3p)}$.

Notice that for discussed expanding wormholes, filled with a scalar
and tachyon fields, the wormhole geometry is given by
\begin{eqnarray}\label{INHOMOGENEOUS FRW915}
ds^2=dt^2-a_0^2 \, t^{2p}  \times \nonumber \\ \left( \frac{dr^2}{1-
\left(\frac{r}{r_0}\right)^{-(1+\omega_r)/\omega_r}}+ r^2 (d\theta^2+sin^2 \theta d \varphi^2)\right). \nonumber \\
\end{eqnarray}
In this case, the energy density of the anisotropic component and
its pressure are given by
\begin{equation}\label{WH FRW915}
\kappa \rho_{_{in}}(t,r)=-\frac{1}{\omega_r r^2_0 a_0^2 \, t^{ 2p}}
\left(\frac{r}{r_0}\right)^{-(1+3\omega_r)/\omega_r}
\end{equation}
and~(\ref{pr pl}) respectively.

\section{Conclusions}

We have developed models for evolving wormholes sustained by two
cosmic fluids: one with homogeneous and isotropic properties and
another inhomogeneous and anisotropic. It is remarkable that the
energy density of the matter threading and sustaining such a
wormhole is the inhomogeneous and anisotropic component, while the
rate of expansion of the evolving wormhole is determined by the
isotropic and homogeneous component. This matter component may be in
principle an ideal barotropic fluid, a scalar field or any other
cosmic fluid satisfying the homogeneity and isotropy requirements.
For the case where the cosmological constant and the isotropic and
homogenous component are absent the inhomogeneous space expands with
constant velocity, and when only the isotropic and homogenous
component is absent the rate of expansion is determined by the
cosmological constant.

In general, we have wormhole universes for $\omega_{r}<-1$ or
$\omega_{r}>0$. The present results generalize our previous
works~\cite{Cataldo1} and~\cite{Cataldo2}. The case when the studied
inhomogeneous geometry represents wormhole configurations expanding
with constant velocity, i.e. $\omega_{r}<-1$ or $\omega_{r}>0$ and
$\rho(t)=\Lambda=0$, was discussed in Ref.~\cite{Cataldo1}, while
the scenarios where the expansion rate is determined by the
cosmological constant, i.e. $\rho(t)=0$, were discussed in
Ref.~\cite{Cataldo2}. If now $\rho(t) \neq 0$ and $\Lambda=0$, the
expansion rate of the wormhole is determined by this isotropic and
homogeneous fluid. It is interesting to note that the results of
Ref.~\cite{Cataldo2} were generalized to the case of evolving
wormholes sustained by a single inhomogeneous and anisotropic fluid
$\varrho(t,r)$, by imposing the generalized equation of state
$\varrho+\alpha P_r+2 \beta P_t=0$, where $\alpha$ and $\beta$ are
constant parameters, and $P_r$ and $P_t$ are the radial and
transverse pressures, respectively~\cite{Riazi}. All particular
wormhole solutions discussed in this Ref. are related to solutions
reported in Ref.~\cite{Cataldo2}.

Note that the wormhole geometries~(\ref{WH FRW})-(\ref{WH FRW915})
far from the throat look like a flat FRW Universe. At first glance,
if the wormhole throat is located outside of the cosmological
horizon of any observer, then he is not in causal contact with the
throat. Thus, for late times, an observer in this wormhole Universe
located too far from the wormhole throat will see the Universe
isotropic and homogeneous and it will be in principle unable for him
to make a decision about whether he lives in a space of constant
curvature or in a space of a wormhole spacetime.

An interesting feature of the discussed here solutions is that one
can consider evolving wormholes with positive total energy density,
i.e. $\rho_{_{total}}=\rho(t)+\rho_{_{in}}(t,r)>0$. We always are
free to consider positive homogeneous and isotropic energy density
$\rho(t)$. In order to have positive $\rho_{_{in}}(t,r)$ we must
choose $\omega_{_r} \leq -1$. In this case the inhomogeneous and
anisotropic matter component sustaining the evolving wormhole may be
considered a generalization of the hypothetical phantom energy used
in cosmology in order to explain accelerated expansion of the
Universe. Effectively, this cosmic phantom source is characterized
by a positive homogeneous energy density, i.e. $\rho_{_{DE}}(t)>0$,
and by an isotropic pressure satisfying $p_{_{DE}}(t)<-
\rho_{_{DE}}(t)$. Clearly for $p_{_{DE}}= \gamma \rho_{_{DE}}$ we
have that $\gamma<-1$. In our case the phantom energy is realized by
the inhomogeneous matter component $\rho_{_{in}}(t,r)>0$ with linear
but highly anisotropic equation of state $p_{_{r}}<-
\rho_{_{in}}<0$, $p_{_l}>0$. Spherically symmetric distribution of
phantom energy, depending only on the radial coordinate $r$, with
such linear equation of state for the radial pressure were
considered in Refs.~\cite{Sushkov}, where the authors constructed
static wormholes sustained by a positive energy density $\rho(r)>0$.

Lastly, let us note that, to our knowledge, the results described in
this article, for evolving lorentzian wormholes, with
energy-momentum tensor associated with a mixture of one isotropic
and homogeneous fluid with an inhomogeneous and anisotropic
component, are firstly reported here. There are in the literature
many works reporting on evolving wormhole solutions to the Einstein
field equations, however, a large number of these papers discuss
dynamic wormhole solutions with a single fluid
source~\cite{EvolvingWH,KarSahdev,Arellano,Riazi}. It is worth to
mention here that a general class of higher evolving dimensional
wormholes sustained by a single fluid was studied in
Ref.~\cite{DeBenedictis}. Most specifically, it was considered a
quasi-static spherically symmetric evolving wormholes, with static
four non-compact dimensions and an arbitrary number of extra
time-dependent compact dimensions. The results of the study show
that the WEC cannot be satisfied at the throat. This is mainly due
to that the matter content is not distributed in the whole space and
is matched to the vacuum. The presence of this matter-vacuum
boundary places restrictions on the time dependence of extra compact
dimensions, consequently implying the violation of the WEC. On the
contrary, as shown in the previous Sec., we can have evolving
wormhole configurations satisfying the WEC.

On the other hand, with respect to wormholes involving a mixture of
two fluids, a dynamical wormhole, filled with a perfect fluid and a
ghost scalar field, is provided in Ref.~\cite{Maeda}. The considered
in this Ref. dynamical wormhole metric (4.53),
\begin{eqnarray*}
ds^2=dt^2-a^2(t) \left( dx^2-(x^2+\tilde{b}^2) d \Omega^2 \right),
\end{eqnarray*}
may be rewritten, by using the transformation $x^2=r^2-\tilde{b}^2$,
as
\begin{eqnarray*}
ds^2=dt^2-a^2(t) \left( \frac{dr^2}{1-\tilde{b}^2/r^2}-r^2 d
\Omega^2 \right),
\end{eqnarray*}
where $\tilde{b}$ is a constant parameter and $a(t)=t/t_0$. Thus the
solution (4.58)-(4.60) of the Ref.~\cite{Maeda} is a particular case
of the discussed in this paper wormhole geometries~(\ref{WH
FRW})-(\ref{C15B}), with $\omega=-1/3$, $\omega_r=1$,
$r_0=\tilde{b}$ and $a_0=1/t_0$. In this case the anisotropic and
inhomogeneous component with $\omega_r=1$ may be identified as a
massless ghost scalar field (note that the energy density~(\ref{WH
FRW15}) becomes negative). The conditions~(\ref{C15A})-(\ref{C15B})
imply the inequalities $a_0^2r_0^2 \geq 1/3$, $a_0^2r_0^2 \geq 1$,
$a_0^2r_0^2 \geq 0$ and $a_0^2r_0^2 \geq 1/2$ respectively. Hence,
for this particular solution the DEC is satisfied in the whole space
for $a_0 r_0 \geq 1$. Notice that the wormhole throat definitions of
Hochberg-Visser~\cite{Hochberg} or Hayward~\cite{Hayward} do not
apply to this evolving wormhole since the whole spacetime is
foliated by trapped surfaces and there is no trapping
horizon~\cite{Maeda}.

As far as we know, there is only one more paper where two-fluid
evolving lorentzian wormholes are considered. In Ref.~\cite{Kim} FRW
models with a traversable wormhole are considered. In this case the
matter content is divided into two parts: the cosmic part $\rho$
that depends on cosmological time only, and the wormhole part
$\rho_w$ that depends on the radial coordinate only. This wormholes
can be finally connected with the particular wormholes
solutions~(\ref{WH FRW})-(\ref{WH FRW15}).

\section{Acknowledgements}
This work was supported by CONICYT through Grants FONDECYT N$^0$
1080530 and 1110230 (MC, SdC), and by Direcci\'on de Investigaci\'on
de la Universidad del B\'\i o-B\'\i o (MC). SdC also was supported
by PUCV grant N$^0$ 123.710/2011.

\end{document}